\documentclass{ws-procs9x6-cpt16}

\def\etal {{\it et al.}}

\begin{document}
\def\gbar{\ensuremath{\bar{g}}}

\title{\vspace{-.45in}\rightline{\footnotesize\rm IIT-CAPP-16-4}\vspace{.25in}
Progress Towards a Muonium Gravity Experiment}

\author{Daniel M.\ Kaplan,$^1$ Klaus Kirch,$^2$ 
Derrick C.\ Mancini,$^1$ James D. Phillips,\\ 
Thomas J.\ Phillips,$^1$ Robert D.\ Reasenberg,$^3$ 
Thomas J.\ Roberts,$^1$ and Jeff Terry$^1$}

\address{$^1$Physics Department, Illinois Institute of Technology,
Chicago, IL 60616, USA}

\address{$^2$Paul Scherrer Institute, Villigen and ETH Z\"{u}rich, Switzerland}

\address{$^3$CASS, UC San Diego, La Jolla, CA 92093, USA}

\begin{abstract}
The gravitational acceleration of antimatter, \gbar, has yet to be directly measured but could change our understanding of gravity, the Universe, and the possibility of a fifth force.  
Three avenues are apparent for such a measurement: antihydrogen, positronium, and muonium, the last requiring a precision atom interferometer and benefiting from a novel 
 muonium beam under development. The
interferometer and its few-picometer  alignment and calibration systems appear to be feasible. With 100\,nm grating pitch, measurements of \gbar\ to  10\%, 1\%, or better  can be envisioned. This could constitute the first gravitational measurement of leptonic matter, of second-generation matter and, possibly, the first measurement of the gravitational acceleration of antimatter.
\end{abstract}

\bodymatter

\phantom{}\vskip10pt\noindent
Despite many years of effort, experiments on antimatter gravity have yet to yield a statistically significant direct measurement. Such studies using antihydrogen and positronium are ongoing. We report here on  progress towards  a measurement using muonium.

{\em Indirect} tests, based on the expected amounts of virtual antimatter in the nuclei of various elements, imply stringent limits on the gravitational acceleration, \gbar, of antimatter on earth: $\gbar/g-1<10^{-7}$.\cite{Alves}\footnote{The extent to which these limits apply to muonium is  far from obvious.} 
A {\em direct} test of the gravitational interaction of antimatter with matter seems desirable on quite general grounds\cite{Nieto-Goldman} and is  of interest whether viewed as a test
of General Relativity
or as a search for a fifth force.
Candidate quantum gravity theories include the possibility of differing  matter--antimatter and matter--matter forces;\cite{Nieto-Goldman}\footnote{For example, suppressed scalar and vector terms may 
cancel in matter--matter interactions, but add in matter--antimatter ones.} 
recent work\cite{Tasson}
on gravity in the SME framework
emphasizes the importance of
second-generation  measurements. 
The short lifetimes of second- and third-generation particles may make
muons the only experimentally accessible  avenue to gravity beyond the 
first generation.  

Although most physicists expect the equivalence principle to hold
for antimatter as well as for matter, theories in which this symmetry
is {\em maximally} violated (i.e., in which antimatter ``falls up'') are
attracting increasing interest\cite{DK1}\footnote{See Ref.\ \refcite{kaplan} for further references as well as a more detailed discussion of our experiment.}
 as potentially solving 
six great mysteries of cosmology (Why is the cosmic microwave background  so isothermal? Why is the Universe so flat? Why are galactic rotation curves flat? What happened to the antimatter? Why does $\Lambda=0$ cosmology give the age of the Universe as younger than the oldest stars, and Type IA supernovae dimmer than predicted?), all without the need for cosmic inflation, dark matter, or dark energy.


We are developing a precision three-grating  atom-beam interferometer for the measurement of \gbar\ using a slow muonium beam  at  Switzerland's Paul Scherrer
Institute (PSI).\cite{Klaus}
The interferometer
can measure the  atoms' gravitational deflection  to a fraction
of a nanometer, determining \gbar\ to a precision of 10\%
of $g$ in a month of beam time (assuming a typical 30\% overall efficiency). Additional  time, intensity, or efficiency could  permit a 
measurement to 1\% or better. 
The RMS statistical precision is estimated as\cite{Kirch}
$\delta g=d/(2\pi C\sqrt{N}\,t^{2}),$
where $C = 0.1$ is the fringe contrast, $N$ the number of events detected, and $t$ the muonium transit time through the interferometer. 
A finer grating pitch $d$ is helpful; we have chosen $d= 100$\,nm  as a compromise between sensitivity and systematic error due to geometry variations over the $\sim$\,cm$^2$ grating area. 
At the anticipated rate of $10^5$ muonium atoms/s incident on the interferometer, 
the statistical measurement precision is about $0.3g$ per $\sqrt{N_d}$, where $N_d$ is the exposure time in days. 

The monoenergetic muonium beam is under development at PSI;\cite{Klaus} a first test using an existing, thermal-muonium beam is also of interest and could potentially provide the first determination of the sign of \gbar. Interferometer development, including Si$_3$N$_4$  grids nanofabricated using $e$-beam lithography and reactive-ion etching at the ANL Center for Nanoscale Materials, is underway using teams of IIT undergraduates  and donated equipment and facility time.\cite{Melanie} A key challenge is the need to translate one grating vertically with at least 10 pm precision in order to scan the interference pattern.  Most recently, using two semiconductor-laser tracking frequency gauges\cite{Thapa-etal} (TFGs) we have demonstrated position measurement to $\approx$\,3 pm, with work ongoing to reduce residual noise.  The 10\,pm requirement is seen (Fig.~1) to imply a need for geometric stability over at least 0.3\,s and calibration with X-rays at least every 1000\,s.

\begin{figure}
\begin{center}
\includegraphics[width=0.7\hsize,trim=0 190 0 68mm,clip]{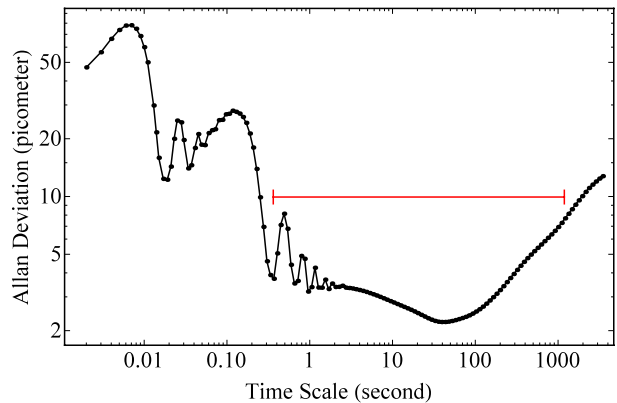}
\end{center}
\caption{Allan deviation vs.\ averaging time obtained at IIT for a two-TFG test showing $\stackrel{<}{_\sim}$\,3\,pm precision with one- to several-second averaging time. The bar at 10\,pm (red online) shows the range of time scales over which the TFG measurement is useful. }
\end{figure}

\section*{Acknowledgments}

We thank the Smithsonian Astrophysical Observatory 
for donation of two TFGs as well as the Physics Department, 
College of Science, BSMP, and IPRO programs at IIT.  
Use of the Center for Nanoscale Materials, an Office of Science user facility, 
was supported by the DoE under contract DE-AC02-06CH11357. 
The development of a suitable muonium beam is supported 
by the Swiss National Science Foundation, grant No.\ 200020\_159754.


\begin{thebibliography}{x}

\bibitem{Alves}
D.S.M.\ Alves, M.\ Jankowiak, and P.\ Saraswat,
arXiv:0907.4110.

\bibitem{Nieto-Goldman}
M.M.\ Nieto and T.\ Goldman,
Phys. Rep. {\bf 205}, 221(1991).

\bibitem{Tasson}
V.A.\ Kosteleck\'y and J.D.\ Tasson,  
Phys. Rev. D {\bf 83}, 016013 (2011).

\bibitem{DK1}
See, e.g., L. Blanchet, 
Class.\ Quant.\ Grav.\ {\bf 24}, 3529 (2007);
L.\ Blanchet and A.\ Le Tiec, 
Phys.\ Rev.\ D {\bf 78}, 024031 (2008); 
A.\ Benoit-L\'evy and G.\ Chardin, 
Astron.\ Astrophys.\ {\bf 537}, A78 (2012);

\bibitem{kaplan}
D.M.\ Kaplan \etal,
arXiv:1601.07222.

\bibitem{Klaus}
K.\ Kirch, these proceedings; 
A.\ Eggenberger, these proceedings.

\bibitem{Kirch}
K.\ Kirch,
arXiv:physics/0702143.

\bibitem{Melanie}
M.\ Dooley \etal,
presented at the 2016 ANL CNM Users' Meeting, May 2016.

\bibitem{Thapa-etal}
R.\ Thapa \etal,
Opt.\ Lett.\ {\bf 36}, 3759 (2011).

\end{thebibliography}
\end{document}